\documentstyle[twoside,fleqn,espcrc2,epsf]{article}

\def\psibar{\overline{\psi}}

\def\d#1/d#2{ {\partial #1\over\partial #2} }

\def\beq{\begin{equation}}
\def\eeq{\end{equation}}
\def\beqa{\begin{eqnarray*}}
\def\eeqa{\end{eqnarray*}}
\def\beqs{\begin{eqnarray}}
\def\eeqs{\end{eqnarray}}
\def\n{\global\advance \eqnumber by 1\eqno(\the\eqnumber)}
\def\puteqno{\global\advance \eqnumber by 1 (\the\eqnumber)}

\title{QCD with zero, two and four flavors of light quarks - results
from {\it {\it QCDSP}} }

\author{Chengzhong Sui\address{Department of Physics, Columbia University,
     New York, NY 10027, USA}\thanks{This work was done in collaboration with
     Igor Arsenin, Dong Chen, Ping Chen, Norman H. Christ,
     Robert Edwards, George Fleming, Alan Gara, Sten Hansen,
     Chulwoo Jung, Adrian Kahler, Anthony Kennedy, Gregory Kilcup, 
     Yubing Luo, Catalin Malureanu, Robert D. Mawhinney,
     John Parsons, James Sexton, Pavlos Vranas and Yuri Zhestkov.  
     This work was supported in part by the 
     Department of Energy. Presented on Lattice '98.}}

\begin{document}

\begin{abstract}
We present the results from full QCD simulations with four flavors of light staggered
dynamical quarks on {\it {\it QCDSP}} supercomputer. Previous results are 
reproduced and the simulation reported here yields new results consistent with our previous runs. 
The hadron spectrum obtained with Wilson valence fermions reported here will
allow us to determine if our earlier conclusions are independent of lattice formalism.

\end{abstract}

\maketitle

\section{INTRODUCTION}
It is well known that the number of light quark flavors $N_f$ governs the order of the QCD phase
transition at finite temperature.
On the other hand the effects of dynamical quarks at zero temperature 
have generally been found to be quite small when zero-flavor (quenched) and two-flavor QCD
simulations are compared. One important question is whether this will continue 
as $N_f$ increases.
In the past few years the Columbia group has done extensive comparison 
studies of zero-flavor, two-flavor and four-flavor QCD at zero 
temperature with the lattice spacing and volume fixed in physical units
determined by the $\rho$ mass. 
What we found was that for four flavors the hadron spectrum and chiral condensate
were significantly different from zero and two flavors: the hadronic
spectrum appeared to be parity doubled and the chiral condensate was
almost four times smaller than that for two flavors\cite{Chen959697}\cite{rdm98}.

Those calculations were done on the Columbia 16 Gflop machine, Fermi256. With the new
generation of Columbia supercomputer, {\it QCDSP}, we have
been able to continue our flavor-dependence study. As a first step we need
to check the reliability of our new hardware and software. We have reproduced
results from several other groups\cite{reproduce} on small lattices,
and we have also tried to reproduce our own earlier results. 
For our previous four-flavor calculations we did two simulations with dynamical
quark mass 0.01 and 0.02. It
is natural to have at least one more data point to be able to do a linear fit 
to the dynamical results. Another interesting question is whether the partial 
chiral symmetry restoration that we
observed is just an artifact of the lattice formalism. 
To investigate this problem
we have also measured the Wilson valence spectrum on these lattices generated
with staggered dynamical quarks.

\section{SIMULATIONS AND RESULTS}

Table~\ref{tab:parameters} lists the run parameters for our 
four-flavor simulations. Run I and III are our previous simulations done on 
Fermi256\cite{rdmJapan} while II and IV are simulations performed on {\it QCDSP}. 
Run IV is a check of results from {\it QCDSP} against previous results from Fermi256.
Run II is a new simulation, adding one
more dynamical data point so that linear fits can be performed.
The four-flavor simulations used an exact hybrid Monte Carlo algorithm,
employing the $\Phi$ hybrid molecular dynamics algorithm of
\cite{Gottlieb87}, with a Monte Carlo accept/reject step.

\begin{table*}[htb]
\caption{Simulation parameters for runs with $N_f = 4$ and $\beta=5.4$ on 
$16^3\times32$ lattice and corresponding results of staggered hadron masses 
with $m_{\rm val} = m_{\rm dyn}$. We have chosen the hadron mass fitting 
range typically from 9 to 16 for the $\pi$ and from 6 to 16 for the other particles.  }

\label{tab:parameters}
\begin{center}
\begin{tabular}{|c|c|c|c|c|}
\hline
\multicolumn{5}{|c|}{$N_f = 4$, $\beta=5.4$} \\\hline
  run               & I  & II & III     & IV  \\\hline
  $m_{\rm dyn}a$        & 0.01  & 0.015 & 0.02    &0.02 \\\hline
  run length            & 4450  & 4390 &2725 & 3980  \\ \hline
  step size             & 0.0078125 &0.005 & 0.01 & 0.005  \\ \hline
  steps per trajectory  & 64    & 100 &50 & 100  \\\hline
  acceptance            & 0.95 & 0.87 &0.99 & 0.91  \\ \hline
  machine               & Fermi256& {\it QCDSP}& Fermi256 & {\it QCDSP} \\ \hline
\hline

$\pi$   & 0.292(5) & 0.320(4) & 0.357(3) & 0.356(2)  \\ \hline
$\rho$  & 0.438(8) & 0.470(5) & 0.499(6) & 0.501(4) \\ \hline
$a_1$   & 0.493(7) & 0.562(10)& 0.617(14)& 0.641(9) \\ \hline
$N$     & 0.690(21)& 0.742(9)& 0.773(10)& 0.781(5) \\ \hline
$N'$    & 0.731(16)& 0.802(15)& 0.872(13)& 0.912(16) \\ \hline
\end{tabular}
\end{center}
\end{table*}

Comparing the hadron masses in run III and IV in Table \ref{tab:parameters},
we find that the hadron spectrum agrees very well statistically between the two
machines. The expectation values of average plaquette and chiral
condensate are also found to be consistent. However there is one discrepancy between the two calculations
which we don't understand. In order to achieve comparable acceptance rate
with the new code we had to use a HMC time step size that was nearly
half as large as that used previously. This is most likely due to
some failure in our earlier program to adhere to the conventions of
\cite{Gottlieb87}. We are trying to resolve this discrepancy.

\pagenumbering{arabic}
\addtocounter{page}{1}

\begin{figure}[hbt]
\epsfxsize=\hsize
\vskip -0.7in
\epsfbox[10 10 587 524]{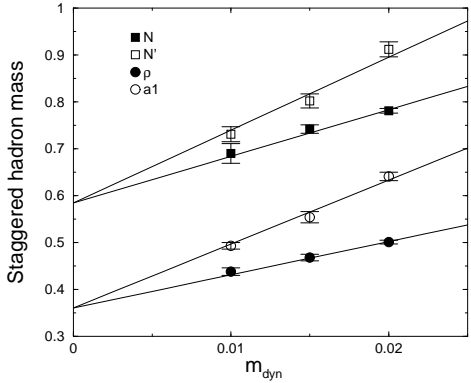}
\vskip -0.4in
\caption{Staggered masses for the $\rho$ and $N$ and their parity
partners vs. dynamical quark mass for $N_f=4$. The fits are three-parameter linear fits to
six data points with common intercept. }
\label{fig:fig1}
\end{figure}

In figure \ref{fig:fig1} we plot the staggered hadron masses of $\rho$ and 
nucleon and their parity partners as a function of the dynamical quark mass 
using results from simulation I, II and IV. To confirm the near degeneracy of 
the parity partners in the zero quark mass limit we have done a 3-parameter fit to the 6 data
points 
for both parity partner pairs, forcing a common intercept in the chiral limit
respectively. 
The fits have $\chi^2/dof = 1.2$ for the N and N' data and $\chi^2/dof
= 0.7$ for the $\rho$ and $a_1$, which strongly support our parity doubling
observation.

\begin{figure}[htb]
\epsfxsize=\hsize
\vskip -0.9in
\epsfbox[10 10 587 524]{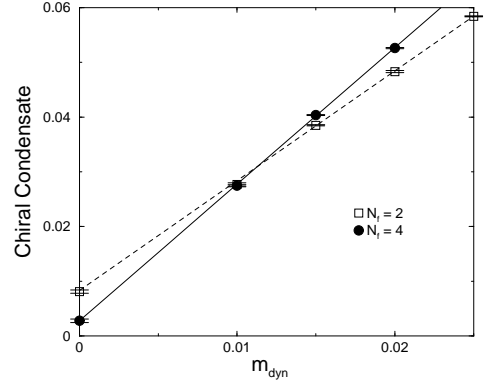}
\caption{$\langle \psibar \psi \rangle$ vs. dynamical quark mass. 
Filled circles are for four flavors and open
squares are for two flavors. The solid line is a linear fit to the 3 four flavor data points and
the dashed line is a fit to the 4 two flavor data points.}
\label{fig:fig2}
\end{figure}

The degeneracy of parity partners implies a restoration of chiral
symmetry breaking. Figure \ref{fig:fig2} shows $\langle \psibar \psi
\rangle$ as a function of the
dynamical quark mass for both $N_f=4$ and $N_f=2$, where the four-flavor
data (solid circle) is taken from run I, II and IV and the two-flavor 
data (open squares) from \cite{rdm98}.
A linear fit to the chiral condensate for four flavors (solid line) yields $\langle \psibar \psi
\rangle = 0.00278(33)$ in the zero-mass limit, which is more than three times
smaller than $\langle \psibar \psi \rangle = 0.00854(17)$ for two flavors.
This indicates significant weakening of chiral symmetry breaking 
as the number of flavors increases, although the four-flavor value is
still about eight standard deviations from zero. The non-vanishing
chiral condensate in the zero-mass limit suggests
that the hadron mass splittings between the parity partners will not
be exactly zero
although they are invisible in figure \ref{fig:fig1} because of the relatively large statistical errors on the hadron masses.

\begin{figure}[htb]
\epsfxsize=\hsize
\epsfbox[10 10 587 524]{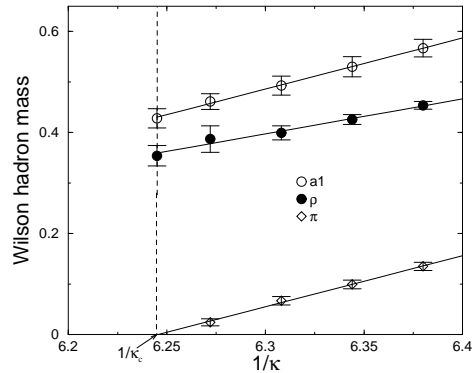}
\caption{$m_{a_1}$, $m_{\rho}$ and $m_{\pi}^2$ vs. $1/\kappa$ in run II
with $N_f = 4$ and $m_{\rm dyn} = 0.015$. The filled circles are masses of $\rho$, open
circles are masses of $a_1$ and diamonds are masses squared of $\pi$. The dashed vertical line shows the
position of $\kappa_c$.}
\label{fig:w_fig}
\end{figure}

In calculating the Wilson valence hadron propagators, we employ the operator $\psibar
\gamma_5 \gamma_3 \psi$ for the $a_1$ and $\psibar \gamma_3 \psi$ for the
$\rho$. Wall sources and Coulomb gauge are also used. In figure \ref{fig:w_fig} the Wilson
valence masses for the $\rho$ and $a_1$ measured
in simulation II, are
plotted as a function of the valence hopping parameter $\kappa$. Linear
fits are performed to both sets of data. The
$\kappa_c$ is determined through the PCAC relation $m_{\pi}^2 = C ( 1/\kappa -
1/\kappa_c)$. We can see that the parity partners split less and less as 
$\kappa$ approaches $\kappa_c$. Similar behavior is also observed in partially
quenched staggered hadron masses from this and earlier runs\cite{rdmJapan}.

In conclusion, the new {\it QCDSP} simulations have reproduced our previous
runs successfully. The new simulation
with $N_f = 4$, $\beta=5.4$ and $m = 0.015$ on {\it QCDSP} produces
results consistent with our previous potential discovery that chiral symmetry breaking
is reduced substantially as $N_f$ increases from 2 to 4. The Wilson valence hadron
spectrum results are consistent with the hypothesis that this result is independent of lattice
formalism, although more simulations are needed.
The possible effects of small volume remain an important concern when 
interpreting our results. To resolve this issue we need to
simulate on lattices of larger spatial
volumes. In fact, larger, $24^3 \times 32$, full QCD simulations are now under way.

\end{document}